\documentstyle[twocolumn,eqsecnum,aps]{revtex}
\newcommand{\vc}[1]{{\bf {#1}}}
\begin{document}
\input{epsf.tex}
%
\draft
\title{Atom-optics hologram in the time domain}
\author{A. V. Soroko\footnote{Electronic address: soroko@hep.by}}
\address{National Centre of Particle and High Energy Physics,
Belarusian State University, \\
Bogdanovich Street 153, Minsk 220040, Belarus}
\maketitle
\begin{abstract}
    The temporal  evolution of  an atomic  wave packet interacting
with object  and reference  electromagnetic waves  is investigated
beyond the weak  perturbation of the  initial state.   It is shown
that  the  diffraction  of   an  ultracold  atomic  beam   by  the
inhomogeneous laser field can be interpreted as if the beam passes
through   a   three-dimensional   hologram,   whose  thickness  is
proportional  to  the  interaction   time.    It  is   found  that the
diffraction efficiency of such a  hologram may reach 100\% and  is
determined by the duration of laser  pulses.  On this basis a
method for reconstruction of the object image with matter waves is
offered.
\end{abstract}
%

\narrowtext

\section{INTRODUCTION}
\label{Introduction}

    The  success of the  last decade in the  field of laser  light cooling
below the recoil limit \cite{VSCPT,Raman} has opened a new chapter
of atom optics, whose objective is to manipulate atomic beams in  a
way  similar  to  conventional  optics  by  exploiting  the   wave
properties of the particles.   Indeed, if the momenta of cooled  atoms
approach those  of  photons, diffraction  effects may manifest
themselves particularly  strongly during  atomic interaction  with
spatially inhomogeneous radiation.  For the corresponding part  of
the  de  Broglie  wave  spectrum,  this  provides the possibility of
supplementing the traditional atom-optics set of elements such  as
mirrors        \cite{amirror},        diffraction         gratings
\cite{Moskowitz,Martin}, or lenses  \cite{alens} with holograms  of
different objects, the conventional optics analogs of which have
been well  known for  several decades  \cite{Gabor}.   Creation of
matter waves with the required amplitude and phase characteristics  is
the main  task assigned  to such  atomic holograms.   Since  these
characteristics are  the same  as those  of the  object wave,  one
obtains a  powerful and  convenient tool  for holographic  imaging
with atoms.   The  latter may  have useful  practical applications
ranging  from  atom  lithography  \cite{atomlithography}  to   the
manufacturing of microstructures or quantum microfabrication.

    One possibility  to make  an atomic  hologram is  to create  a
mechanical  mask  with  appropriate  transparency for the incident
atomic beam (the analog of a two-dimensional optical hologram).   Such
a hologram has the advantage  of being permanent.  However,  up to
now only masks with binary  transparency have been prepared.   For
example, in  the experiment  \cite{Moringa} the  mask was  written
onto  a  thin  silicon  nitride  membrane  and  allowed for either
complete or vanishing transmission of  the beam at a given  point.
Since gradually varying transmission  of the beam is  required for
the  correct  holographic  storage  of  information,  this reduces
resolution in the reconstructed image.

    A very interesting proposal has recently been reported in
Ref.\  \cite{ahbose},  where  the  authors  regard a  Bose-Einstein
condensate  (BEC)  as  the  registration  medium  for  the   atomic
hologram.  In the suggested method the information issuing from an
object  is  encoded  into  the  condensate  in the form of density
modulations by using  the object and  reference laser beams  which
form the writing optical potential.  The reconstruction of  matter
wave arises due to s-wave scattering of the reading-beam atoms  on
condensate inhomogeneities.   This  proposal illustrates  the wide
potential applicability of the BEC,  which, since it was  realized
experimentally \cite{BEC},  is now  available almost  routinely in
several laboratories.

    In a previous  paper \cite{creation} we  have shown that  an
atomic hologram may also  be constructed under certain  conditions
as a superposition of  reference and object electromagnetic  waves,
which  is  common  for  optical  holography.   The creation of the
intended  matter  wave  occurs  when  an ultracold  atomic  beam   is
diffracted  from  this   hologram,  which  in   fact  is  just   an
inhomogeneous light field.   The main  advantages of the  proposed
scheme are its  simplicity because of  bypassing the  recording
process  and,  as  a  consequence,  the absence  of aberrations in the
stored information.   In this sense  our approach is  close to the
non holographic scheme of wave front engineering  \cite{Olshanii},
in  which  the  center-of-mass  wave  function  of  an atom can be
arbitrarily shaped by means of a sequence of suitably formed laser
pulses.

    The main assumption employed in our holographic scheme is  the
linear  response   of  the   atomic  system   to  the  laser-field
inhomogeneity.  It requires, in particular, that the  perturbation
of the incident atomic  beam is weak, and  sets an upper limit  on
the  object   wave  amplitude   [see  Eq.\   (50)  in   Ref.\
\cite{creation}].  As a result,  only a small portion of  atoms in
the beam can  be transferred into  the reconstructed matter  wave.
Operating in the linear-response regime decreases the  diffraction
efficiency of an atomic hologram, i.e., the ratio of the  intensity
of diffracted atomic waves to  the intensity of the reading  beam,
which may be crucial for practical applications.  In  conventional
optical holography such a situation corresponds to the kinematical
regime of information recording \cite{Gabor}.  On the other  hand,
the  coupled  wave  theory  of  Kogelnik  \cite{Kogelnik}, and the
theories     based     on     the     dynamical      approximation
\cite{Ewald,Sidorovich,Oberthaler}, provide a recipe for creating
a hologram  with high  (up to  100\%) diffraction  efficiency.  To
achieve  this  goal  it  is  necessary  to  control,  among  other
parameters,
\begin{figure}
\epsfbox{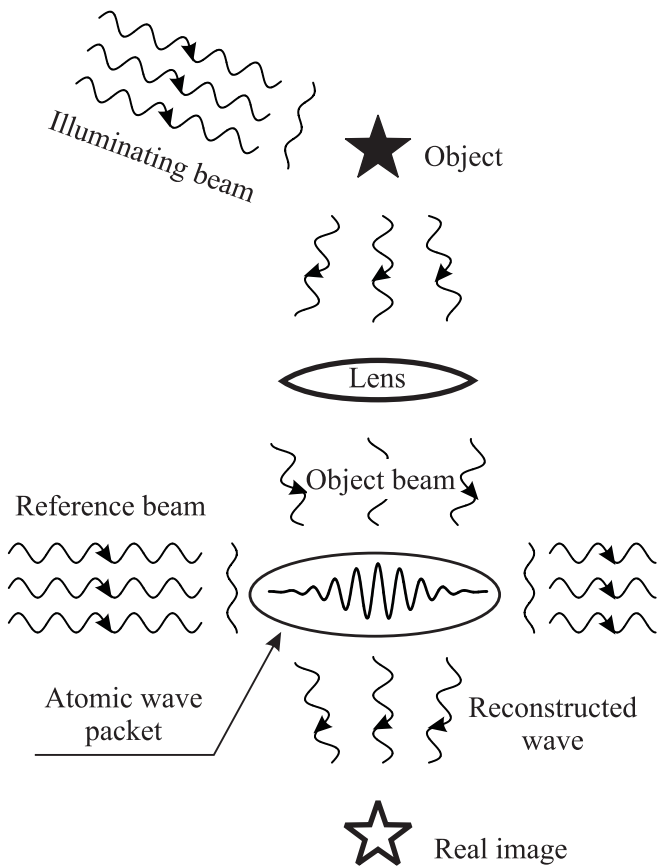}
\vspace{0.5cm}
\caption{
    Typical layout design of  laser beams and matter  wave packets
suitable for atomic holography.
}
\label{fig1}
\end{figure}
the  thickness  of  the  hologram.     Unfortunately,
thickness  controlling  is  difficult   in  a  scheme  of   atom
holography  without  a registration  medium  like  the  one in Ref.\
\cite{creation} (see Fig.\ \ref{fig1} for details).

    The purpose of the present paper is to suggest an  approach
for  creation  of  atom  optics  holograms that will combine the
advantages   of   our   previous   scheme  with  high  diffraction
efficiencies.  We will show that high diffraction efficiencies can
be realized if one restricts the extent of the atomic hologram  in
the time domain rather than in space.  The hologram will thus work
in a pulsed regime pumping atoms from the beam or from the initial
wave packet into the reconstructed wave.  Note that the  suggested
regime  is  well  compatible   with  Raman  cooling   methods
\cite{Raman}  (including  laser  cooling  below  the gravitational
limit \cite{cooling}) and the recent realization of an atom  laser
\cite{alaser}, which are  capable of repeatedly  reproducing the coherent
or  almost  coherent  atomic  wave  packets  necessary  for actual
implementation of a reading beam.

    Another  important   prerequisite  for   successful  wave front
reconstruction  with  massive  particles  concerns  the  need   to
compensate   for   the   potentially   detrimental   influence  of
gravitational effects.  Fortunately, the bulk of the atoms possess a
magnetic  moment,  and  all   one  has  to  do  is   use  the
Stern-Gerlach  effect.    Superimposing  the  weakly inhomogeneous
magnetic  field  onto  the  path  of  prepolarized  particles  and
appropriately  adjusting  the  field  gradient,  it is possible to
suspend the  ground-state atoms  everywhere except in the region  of
interaction with  radiation.   But if  the laser  frequency is far
from all atomic transitions,  the contribution to the  total force
induced  by  spatially  dependent  shifts  of the Zeeman levels is
negligible.  Under this condition, atoms move like free  particles
that are affected only by the electromagnetic waves.

    In  Sec.\  \ref{sec2}  we   specify  the  magnetic  field   to
compensate for the  gravity effects and  derive an equation  that
describes the dynamics  of  ground-state  atoms  interacting with the
object and reference beams.  An approximate solution of this equation
is  found   without  assuming   linear  response  to  laser-field
inhomogeneity,  and  its  domain  of  validity is determined.  For
reasonable  experimental   conditions  the   solution  admits   an
atom-optics interpretation, which is presented in Sec.\ \ref{sec3},
namely, the inhomogeneous laser radiation is shown to behave  like
a three-dimensional  hologram with  respect to  the impinging wave
packets.  A numerical simulation of such a hologram created with a
$31$-mode  object  beam  is  carried  out,  and  high  diffraction
efficiency  is  explicitly   demonstrated.    Section   \ref{sec4}
concludes with  a summary  of the  results.   Certain mathematical
details concerning derivation of the basic formulas are  relegated
to the Appendixes.

\section{BASIC FORMULAS}
\label{sec2}

\subsection{Compensation for gravity}
\label{sec2.1}

    To be specific, let us consider an atom with a $J=\case{1}{2}$
to $J=\case{3}{2}$ transition, e.g.,  sodium or cesium.  A magnetic
field $\vc{B}(\vc{r})$ is applied  to compensate for gravity.   It
is supposed to contain a homogeneous component $\vc{B}_0$ directed
along the gravity acceleration $\vc{B}_0 \uparrow\uparrow \vc{g}$.
The remaining inhomogeneous part of the field  $\vc{B}_{1}(\vc{r})
=  \vc{B}(\vc{r})  -  \vc{B}_0$  should  be small compared to this
component,
\begin{equation}
|\vc{B}_{1}(\vc{r})| \ll B_0 = |\vc{B}_0|.
\label{1.0}
\end{equation}
    As we will see below, to fulfil this condition it is necessary
to take $B_0$ in the range $10^3 - 10^4$ G. In practice such  a
field is strong enough to induce Zeeman shifts which  considerably
exceed the  hyperfine splitting  intervals $\sim  \hbar\omega_{\rm
HFS}$ (but not the multiplet ones).  Therefore an internal  atomic
eigenstate $|J,I,M_J,m_I \rangle$ may  be parametrized by the  set
of  quantum  numbers  consisting  of  the  angular  momenta of the
electronic shell $J$ and nucleus $I$, and their local  projections
onto the direction  of the magnetic  field, $M_J$ and  $m_I$.  The
corresponding  energy  eigenvalue  is  determined  not only by the
multiplet level $E_J$ but also by the magnetic field $B(\vc{r})  =
|\vc{B}(\vc{r})|$ and therefore is spatially dependent,
\begin{eqnarray}
E_{|J,I,M_J,m_I\rangle}(\vc{r}) = &&
E_J + a M_J m_I \nonumber \\
&& + (\mu_B g_L M_J - \mu_{\rm nuc} m_I) B(\vc{r}),
\label{eq2}
\end{eqnarray}
    where  $a$  is  the  hyperfine  coupling  constant  ($a\propto
\hbar\omega_{\rm  HFS}$,  e.g.,  for  Na  $a/\hbar  = 885.8$ MHz),
$g_{L}$ denotes the Land\'{e}  factor, and $\mu_{\rm nuc}$  is the
nuclear magnetic  moment.   Because of  the condition (\ref{1.0}),
such  a  spatial  dependence,  however,  mainly  arises  from  the
longitudinal   [$B^{\parallel}_{1}(\vc{r})   =   \vc{B}_0    \cdot
\vc{B}_{1}(\vc{r})    /B_0$],    rather    than   the   transverse
[$\vc{B}^{\perp}_{1}(\vc{r})$]    component    of    the    vector
$\vc{B}_{1}(\vc{r})$,  provided  that  the  components are defined
relative to $\vc{B}_0$.  This is evident from the expression
\begin{eqnarray}
B(\vc{r}) = &&
\sqrt{\left[B_0 + B^{\parallel}_{1}(\vc{r})\right]^2
+ \left[\vc{B}^{\perp}_{1}(\vc{r})\right]^2} \nonumber \\
&& \simeq B_0 + B^{\parallel}_{1}(\vc{r}) +
\left[\vc{B}^{\perp}_{1}(\vc{r})\right]^2/(2 B_0),
\label{1.3}
\end{eqnarray}
    where  the  term  containing  $\vc{B}^{\perp}_{1}(\vc{r})$  is
small  and  can  be  neglected.    Consequently,  by adjusting the
gradient of the field $B^{\parallel}_{1}(\vc{r})$ one can  achieve
translation   invariance   of   the   ground  state  $|g\rangle  =
|1/2,I,-1/2,I \rangle$  (or another  state with  $J=1/2$) in three
dimensions:
\begin{equation}
E_{|g\rangle}(\vc{r}) - M\vc{g}\cdot\vc{r}={\rm const}.
\label{eq3}
\end{equation}
    For example, to  balance the gravitational  force in this  way
for  sodium  it  is   necessary  to  create  a   gradient  $\nabla
B^{\parallel}_{1}(\vc{r})  =  b_1  \vc{g}/|\vc{g}|$,  where $b_1 =
-4.033$ G$/$cm.   This condition does  not contradict the  Maxwell
equation $\nabla \cdot \vc{B}_{1}(\vc{r}) = 0$, because  variation
of $\vc{B}^{\perp}_{1}(\vc{r})$ is not restricted.  Note also that
the  choice  $B_0  =   10^3  -  10^4$  G   maintains  condition
(\ref{1.0}) very well  within a spatial  region of size $\sim
10$ cm.

    All the  other levels  are affected  by the  residual external
potential.    In  particular,  the  force $\vc{f}_e$ acting on the
atoms  in  the  excited  state,  e.g.,  $|e\rangle = |3/2,I,-3/2,I
\rangle$, may be estimated from Eqs.\ (\ref{eq2}) and  (\ref{eq3})
as $|\vc{f}_e| \sim Mg$.

\subsection{Dynamics of the ground-state atoms}
\label{sec2.2}

    In  our  scheme,  we  use  pulses  of laser light at frequency
$\omega$ which is roughly  tuned to the $|g\rangle  \to |e\rangle$
transition.    If  the  typical  size  $2L$  of the atomic  sample is
restricted  by  the  condition  $L  \ll  a/(Mg)$,  one  may regard
$E_{|e\rangle} (\vc{r})$ as the excited level that is closest  to
resonance  within  the  whole  interaction  domain.    Indeed, the
maximal spatial shift of the level $\sim MgL$ induced by the force
$\vc{f}_e$ appears to  be much less  than the hyperfine  splitting
intervals  ($MgL  \ll  a  \sim  \hbar  \omega_{\rm HFS}$), and the
hierarchy of detunings is  retained.  Therefore an  atom initially
in the $|g\rangle$ state behaves as a two-level system with respect to
the processes with stimulated emission of photons.

    The atom moves inside a superposition of the reference and the
object  beams  during  the  whole  laser  pulse.    Each  beam  is
represented   as   a   discrete   sum   of   plane   monochromatic
electromagnetic  waves.    In  particular,  we  use  the following
decomposition of the electric field in the object beam:
\begin{equation}
\vc{E}_{s}(\vc{r},t)=\sum_{m\geq 1}
\vc{E}_{m}
\exp(i\vc{k}_{m}\cdot\vc{r} - i\omega t) + {\rm c.c.},
\label{2.2.1}
\end{equation}
    where  $\vc{E}_{m}$  and  $\vc{k}_{m}$  stand  for the complex
amplitude of the mode $m$ and its wave vector, respectively.   Such
an approach does not restrict the generality of our consideration,
because the expression (\ref{2.2.1})  must well describe the  real
laser field only in the atom-laser interaction region.  Evidently,
the latter requirement can  always be satisfied by  decreasing the
minimal angle between the mode wave vectors.  In this case we  can
also regard the  reference beam as  a single mode  (with the index
$m=0$),
\begin{equation}
\vc{E}_{r}(\vc{r},t)=
\vc{E}_{0}
\exp(i\vc{k}_{0}\cdot\vc{r} - i\omega t) + {\rm c.c.},
\label{2.2.2}
\end{equation}
which is a typical arrangement for optical holography.

    Since  the  atomic  dipole  momentum  operator $\hat\vc{d}$ is
diagonal in quantum numbers  $I$ and $m_I$, the  transitions that
change $m_I$ are allowed only due to hyperfine interaction.  As  a
consequence, the  excited state  $|e\rangle$ decays  to the  lower
ones preferentially in the channel $|e\rangle \to |g\rangle$ (with
the rate $\gamma$).  This  circumstance makes it possible to  deal
with an  atom as  a two-level  system even  if spontaneous  photon
emission takes place.  However, to simplify the consideration, the
coherent  scattering  processes   are  assumed  to   dominate  the
spontaneous emission,  i.e., the  regime $|\Delta|  \gg \gamma$ is
maintained   \cite{Moler,Korsunsky},   where   $\Delta=  \omega  +
[E_{|g\rangle}(0) - E_{|e\rangle}(0)]/\hbar$ is the detuning  from
resonance  in  the  center  of  the  atom-laser interaction region
($\vc{r}=0$).   Under such  a condition  the one-particle  density
matrix  in  the  momentum  representation  \cite{Kazantsev} has an
obvious time evolution,
\begin{eqnarray}
\rho_{ab}(\vc{p}_1,\vc{p}_2,t)= &&
\int d \vc{p}'_1 \int d \vc{p}'_2
\sum_{a'b'} G_{aa'}(\vc{p}_1,\vc{p}'_1,t)
\nonumber\\
&& \times G^{*}_{bb'}(\vc{p}_2,\vc{p}'_2,t)
\rho_{a'b'}(\vc{p}'_1,\vc{p}'_2,t=0).
\label{2.2.3}
\end{eqnarray}
    Here  indices  $a,b  \dots$  span  the  internal atomic states
($e,g$), and $G_{aa'}(\vc{p}_1,\vc{p}'_1,t)$ is the Green function
of  the  two-component  Schr\"{o}dinger  equation describing atomic
dynamics   during   the   $|g\rangle   \leftrightarrow  |e\rangle$
transitions [see Eq.\ (\ref{a.1}) for the details].

    For the situation at hand,  the upper electronic state can  be
adiabatically   eliminated   from   consideration   (see  Appendix
\ref{appenda}) provided that the detuning $\Delta$ is large enough
\cite{Martin,Moler},
\begin{equation}
|\Delta| \gg |\Omega_{m}|,|\vc{f}_e|L/\hbar,
\label{2.2.4}
\end{equation}
    where     $\Omega_{m}     =     \langle      e|\hat\vc{d}\cdot
\vc{E}_{m}|g\rangle/ \hbar$  is the  Rabi frequency  of mode  $m$.
As a result, the dynamics of the ground atomic state is completely
determined by  the equation  for the  center-of-mass wave function
$\psi_{g}(\vc{p},t)$,
\begin{eqnarray}
i\frac{\partial}{\partial t} \psi_{g}(\vc{p},t) = &&
[w(\vc{p})+\Delta+f_0]\psi_{g}(\vc{p},t) \nonumber\\
&&+\sum_{m\geq 1}\biggl(
\sum_{\scriptstyle n\geq 1 \atop \scriptstyle n\neq m}
f_{mn} \psi_{g}[\vc{p}-\hbar(\vc{k}_{m}-\vc{k}_{n}),t]
\nonumber\\
&&+g_{m} \psi_{g}[\vc{p}-\hbar(\vc{k}_{m}-\vc{k}_{0}),t]
\nonumber\\
&&+g^{*}_{m} \psi_{g}[\vc{p}+\hbar(\vc{k}_{m}-\vc{k}_{0}),t]
\biggr),
\label{2.2.5}
\end{eqnarray}
    where  $w(\vc{p})$  denotes  the  kinetic  energy (in units of
$\hbar$), and $f_0$, $f_{mn}$, and $g_{m}$ stand for the effective
Rabi frequencies, introduced by Eqs.\ (\ref{a.3}).

\subsection{Evolution of wave packets}
\label{sec2.3}

    It is known  from the theory  of thick optical  holograms that
reconstruction of the original (conjugate) object wave arises only
if the  reading beam  is directed  along (opposite)  the reference
wave and has  the same wavelength.   In analogy  with conventional
optics,  let  us  consider  for  definiteness  the evolution of an
atomic wave packet whose spectrum is initially concentrated around
the mean momentum  of photons in  the reference beam.   In such  a
case one  can expect  creation of  a matter  wave similar to the
forward  object  wave.    Therefore  it  is convenient to seek the
solution  of  Eq.\  (\ref{2.2.5})   as  a  sum  of   wave  packets
approaching the plane modes of the hologram \cite{Sidorovich},
\begin{equation}
\psi_{g}(\vc{p},t) =
\sum_{m\geq 0} \psi_{m}(\vc{p}-\hbar\vc{k}_{m},t).
\label{2.3.1}
\end{equation}

    Initially  there  are  no  wave  packets  corresponding to the
object beam, so that
\begin{equation}
\psi_{m}(\vc{p},t=0) = 0, \quad m\geq 1,
\label{2.3.2}
\end{equation}
and as a consequence,
\begin{equation}
\psi_{0}(\vc{p}-\hbar\vc{k}_{0},t=0) =
\psi_{g}(\vc{p},t=0)
\equiv \psi_{g}(\vc{p}).
\label{2.3.3}
\end{equation}
    Population of these atomic motional states ($m\geq 1$)  arises
due  to  coupling  with  $\psi_{0}(\vc{p},t)$,  the  wave   packet
corresponding to the reference beam,
\begin{equation}
i\frac{\partial}{\partial t} \psi_{m}(\vc{p},t) =
w_{m}(\vc{p})\psi_{m}(\vc{p},t)
+g_{m} \psi_{0}(\vc{p},t),
\label{2.3.4}
\end{equation}
where
\begin{equation}
w_{m}(\vc{p}) = w(\vc{p}+\hbar\vc{k}_{m})
+\Delta + f_0.
\label{2.3.5}
\end{equation}
Depletion of the state with $m=0$ is governed by the equation
\begin{equation}
i\frac{\partial}{\partial t} \psi_{0}(\vc{p},t) =
w_{0}(\vc{p})\psi_{0}(\vc{p},t)
+\sum_{m\geq 1} g^{*}_{m} \psi_{m}(\vc{p},t)
+\chi(\vc{p},t),
\label{2.3.6}
\end{equation}
    which one can  obtain after substituting  Eqs.\ (\ref{2.3.1}) and
(\ref{2.3.4})  into  Eq.\  (\ref{2.2.5}).

    Equation (\ref{2.2.5}) is thus reduced to the system of  equations
(\ref{2.3.4}),(\ref{2.3.6}).  The advantage of this step  becomes
obvious  after  making  the  following  self-consistent assumption
about the momentum  spectrum of  $\psi_{m}(\vc{p},t)$, $m\geq  0$, the
validity  of  which  was  verified  for the two-mode  case  in  Ref.\
\cite{cooling}.    We   will  suppose   below  that  all
nonvanishing   functions   have   narrow   distributions   around
$\vc{p}=0$ and, as a result, do not overlap in the expression  for
$\chi(\vc{p},t)$,
\begin{eqnarray}
\chi(\vc{p},t) = &&
\sum_{m\geq 1}\biggl(
\sum_{n\geq 1}
g_{n} \psi_{m}[\vc{p}+
\hbar(2\vc{k}_{0}-\vc{k}_{n}-\vc{k}_{m}),t]
\nonumber\\
&&+\sum_{\scriptstyle n\geq 0 \atop \scriptstyle n\neq m}
g^{*}_{m} \psi_{n}[\vc{p}+\hbar(\vc{k}_{m}-\vc{k}_{n}),t]
\nonumber\\
&&+\sum_{\scriptstyle n\geq 1 \atop \scriptstyle n\neq m}
\sum_{l\geq 0} f_{mn}
\nonumber\\
&&\times \psi_{l}[\vc{p}
+\hbar(\vc{k}_{0}-\vc{k}_{l}-\vc{k}_{m}+\vc{k}_{n}),t]
\biggr).
\label{2.3.7}
\end{eqnarray}
    Under  these  conditions  different  parts of $\chi(\vc{p},t)$
give incoherent contributions,  which are small  at low $g_m$  and
can  be  taken  into   account  perturbatively.    In   zero-order
approximation  one  omits  $\chi(\vc{p},t)$  so  that  the  system
(\ref{2.3.4}),(\ref{2.3.6})  becomes  homomorphic  with the rate
equations  describing  an  $(m+1)$-level  atom.    Note  that  the
stationary  solutions  of  this  truncated system exactly coincide
with   eigenmodes   of   the   corresponding   optical    hologram
\cite{Sidorovich}.

    To  make  further  progress  it  is  convenient to perform the
Laplace transformation ($m\geq 0$),
\begin{equation}
\psi_{m}(\vc{p},\lambda) =
\int_{0}^{\infty} dt e^{-\lambda t}
\psi_{m}(\vc{p},t)
\label{2.3.8}
\end{equation}
    with  the  initial  conditions  (\ref{2.3.2}) and  (\ref{2.3.3}).
Then the equations for the  Laplace transforms will allow an  easy
zero-order solution,
\begin{mathletters}
\begin{equation}
\psi^{(0)}_{0}(\vc{p},\lambda) =
\frac{-i}{T(\vc{p},\lambda)}
\psi_{g}(\vc{p}+\hbar\vc{k}_{0}),
\label{2.3.9a}
\end{equation}
\begin{equation}
\psi^{(0)}_{m}(\vc{p},\lambda) =
\frac{-g_m}{w_m(\vc{p}) -i\lambda}
\psi^{(0)}_{0}(\vc{p},\lambda), \quad m\geq 1,
\label{2.3.9b}
\end{equation}
\label{2.3.9}
\end{mathletters}
where
\begin{equation}
T(\vc{p},\lambda)=
w_0(\vc{p}) -i\lambda+
\sum_{m\geq 1}
\frac{-|g_m|^2}{w_m(\vc{p}) -i\lambda}.
\label{2.3.10}
\end{equation}

Similarly, the next iteration reproduces the first-order solution,
\begin{mathletters}
\begin{equation}
\psi^{(1)}_{0}(\vc{p},\lambda) =
\psi^{(0)}_{0}(\vc{p},\lambda)+
\frac{-\chi^{(0)}(\vc{p},\lambda)}
{T(\vc{p},\lambda)},
\label{2.3.11a}
\end{equation}
\begin{equation}
\psi^{(1)}_{m}(\vc{p},\lambda) =
\frac{-g_m}{w_m(\vc{p}) -i\lambda}
\psi^{(1)}_{0}(\vc{p},\lambda), \quad m\geq 1,
\label{2.3.11b}
\end{equation}
\label{2.3.11}
\end{mathletters}
\begin{figure}
\epsfbox{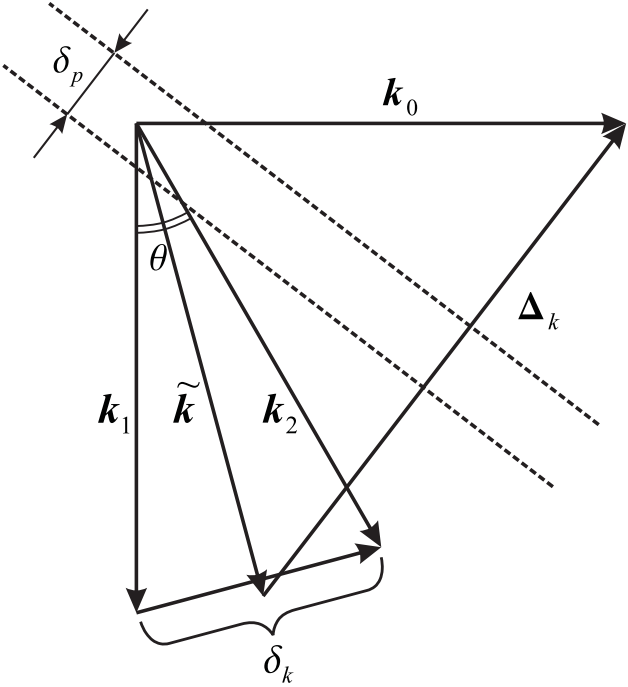}
\vspace{0.5cm}
\caption{
    Definitions  of  the  values  $\vc{\Delta}_k$, $\delta_k$, and
$\delta_p$ in the simplest case of a two-mode object wave.
}
\label{fig2}
\end{figure}
    where  $\chi^{(0)}(\vc{p},\lambda)$   is  obtained   from  the
expression   (\ref{2.3.7})   after   making   the    substitutions
$\psi_{m}(\vc{p},t)  \to  \psi^{(0)}_{m}(\vc{p},\lambda)$,  $m\geq
0$.    In  principle,  we  can  get  the solution with any desired
accuracy by repeating the iterations but it will be sufficient  to
restrict ourselves to the first-order formulas in what follows.

    The time-dependent wave functions are obtained as the  inverse
Laplace transforms of $\psi_{m}(\vc{p},\lambda)$ according to  the
Mellin formula
\begin{equation}
\psi_{m}(\vc{p},t) = 2\pi i
\int_{\epsilon-i\infty}^{\epsilon+i\infty}
d\lambda  e^{\lambda t}
\psi_{m}(\vc{p},\lambda),
\label{2.3.12}
\quad \epsilon > 0.
\end{equation}
    Finally, using Eqs.\ (\ref{2.3.11}) and (\ref{2.3.12}) one can
easily       obtain        the       ground-state        component
$G_{gg}(\vc{p},\vc{p}',t)$ of the Green function appearing in  the
formula (\ref{2.2.3}).  The corresponding expression is derived in
Appendix \ref{appendb}.

\subsection{Validity of assumptions}
\label{sec2.4}

    Let us first check that the zero-order solution  (\ref{2.3.9})
indeed has a narrow momentum spectrum around $\vc{p}=0$,  provided
the initial conditions are chosen properly and the effective Rabi
frequency $g_m$  is small  enough.   We may  restrict ourselves to
examination of  the region  ${\cal D}  = \{\vc{p}:\; [w_m(\vc{p})-
w_0(\vc{p})]^2  \lesssim  |g_m|^2,\;  \forall  m\}$, where all the
functions  in  the  truncated  system  of equations (\ref{2.3.4}),(\ref{2.3.6})
can  come  into  resonance.    In  this  region the
equations admit  an approximate  analytical solution  presented in
Appendix \ref{appendc}.

    The main feature of the near-resonance solution (\ref{c.1}) is
that  the  initial  atomic  wave  packet  transforms into motional
states with $m\geq 1$ at a time $\tau_n$ (time of the $n\pi$ pulse
\cite{Korsunsky})
\begin{equation}
\tau_n=\frac{\pi}{2 g_{\Sigma}}(2n-1),
\label{2.4.1}
\quad n\in \cal{N}.
\end{equation}
    Here  $g_{\Sigma}$  stands  for  the  overall  effective  Rabi
frequency defined  in  Eq.\ (\ref{c.6}).   This  transition is
velocity selective and is most efficient when the Bragg  resonance
condition  $\vc{p}\cdot  \vc{\Delta}_k  =  0$  is satisfied, where
$\vc{\Delta}_k$  denotes  the  typical  difference  between   wave
vectors  in  the  object  and  the  reference  beams  (cf.   Ref.\
\cite{Moler}).  The width of the peak in the momentum distribution
along the direction of  vector $\vc{\Delta}_k$ (the interval  from
the maximum to the first minimum) depends on the interaction time,
and for $t \leq 2\tau_1$ is
\begin{equation}
\delta_p(t) =
\frac{2 M g_{\Sigma}}{\Delta_k}
\sqrt{4\left(\frac{\tau_1}{t}\right) -1}.
\label{2.4.2}
\end{equation}
    For  a  given  value  $\Delta_k= |\vc{\Delta}_k|$ it decreases
with  $g_{\Sigma}$.    Therefore  the  smaller  the effective Rabi
frequencies   $g_m$   the   narrower   the  momentum  spectrum  of
$\psi^{(0)}_{m}(\vc{p},t)$.

    To  prevent  all  nonvanishing  functions comprising the term
$\chi(\vc{p},t)$  from  overlapping  in  momentum space their
spectra  must  be   concentrated  within  the   domain  $|\vc{p}|<
\hbar\delta_k$ at $t \sim \tau_1$, where
\begin{equation}
\delta_k=\min_{m,n \geq 0}
|\vc{k}_m - \vc{k}_n|
\label{2.4.3}
\end{equation}
    is the minimal distance between different wave vectors of  the
laser beams (see  Fig.\ \ref{fig2}).   Since the spectral  extent
along the direction of vector $\vc{\Delta}_k$ is characterized  by
$\delta_p(t)$, we immediately get the condition
\begin{equation}
\delta_p(\tau_1) \ll \hbar\delta_k.
\label{2.4.4}
\end{equation}
    In  agreement  with  Eq.\  (\ref{2.4.2}) it sets an upper
limit on the overall effective Rabi frequency,
\begin{equation}
g_{\Sigma} \ll
\hbar\delta_k\Delta_k/(2\sqrt{3} M).
\label{2.4.5}
\end{equation}

    In the transverse direction the  spectra are the same as  that
of the initial   wave  packet   $\psi_{g}(\vc{p}+  \hbar\vc{k}_{0})$.
Therefore one must impose another condition,
\begin{equation}
|(\vc{p}'-\hbar\vc{k}_{0})\times \vc{\Delta}_k|
< \hbar\delta_k \Delta_k,
\label{2.4.6}
\end{equation}
    which restricts allowed values  of $\vc{p}'$ in the  domain of
the Green function $G_{gg}(\vc{p},\vc{p}',t)$.

    When   the   inequalities   (\ref{2.4.4}) and (\ref{2.4.6})  are
satisfied,  the  main   correction  to  the   zero-order  solution
$\psi^{(0)}_{m}(\vc{p},t)$  caused  by  the  term $\chi(\vc{p},t)$
arises outside the near-resonance region $\cal{D}$ and depends  on
the  geometry  of  laser  beams.    So  for  $t\sim \tau_1$
and a two-dimensional (2D)
holographic setup  like that  in  Fig.\ \ref{fig2}  (i.e., all
$\vc{k}_{m}$ are coplanar
\begin{figure}
\epsfbox{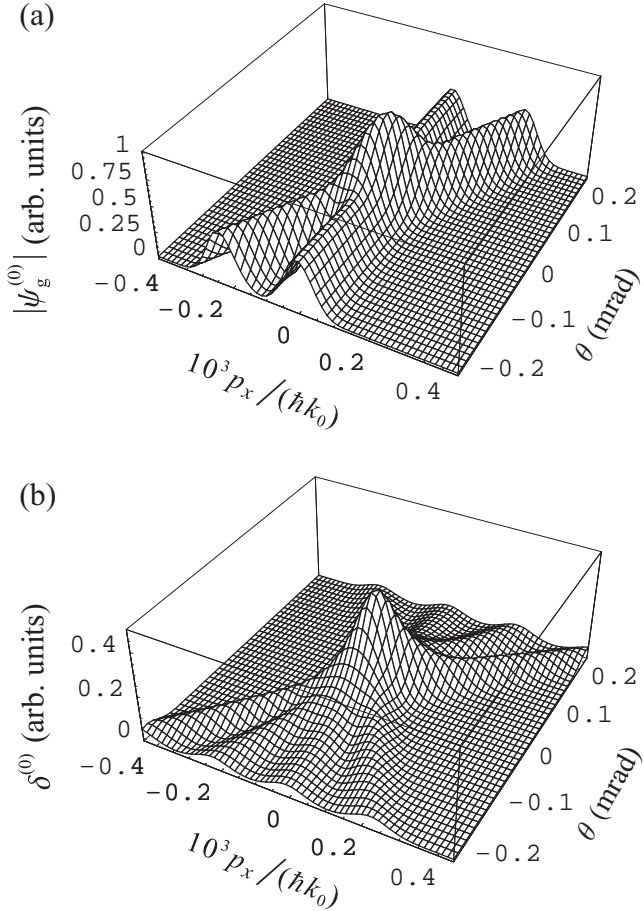}
\vspace{0.5cm}
\caption[Figure 3]{
    Absolute values of zero-order solution $|\psi^{(0)}_{g}|$  (a)
and first-order correction to  it $\delta^{(0)}$ (b) as  functions
of the  momentum component  $p_x$ and  the angle  $\theta$ between
$\vc{k}_{1}$ and $\vc{k}_{2}$.   The arbitrary units scale  is the
same in  both plots.   The  remaining components  of $\vc{p}$  are
fixed as follows:  $p_y= \hbar k_0$ and $p_z= 0$.  The geometry of
the laser beams is chosen as  in Fig.\ \ref{fig2}.   The effective
Rabi frequencies $g_1= g_2= 10$ Hz, $f_{12}= 0.1$ Hz.
}
\label{fig3}
\end{figure}
vectors) the relative correction has the
order        of        magnitude        $\varepsilon_{r}         =
\delta_p(\tau_1)/(\hbar\delta_k) \ll 1$, as follows from  Eqs.\
(\ref{2.3.7}) and  (\ref{c.1.b}).   To obtain  this estimate one
must apply the standard holographic restriction on the intensities
of  laser  beams  $|\vc{E}_{0}|^2\gg  |\vc{E}_{m}|^2$, $m \geq 1$,
which leads to  the inequality $|g_{m}|  \gg |f_{mn}|$.   Then the
third term  in the  expression (\ref{2.3.7})  should be discarded,
because it is proportional to $f_{mn}$ and, consequently, is  much
less than the first and second ones ($\propto |g_{m}|$).

    To illustrate the consistency of  our approach in the case  of a
two-mode  object  wave  let  us  consider  absolute  values of the
zero-order   solution   $|\psi^{(0)}_{g}|$   and  the  first-order
correction to it, $\delta^{(0)}=|\psi^{(1)}_{g}- \psi^{(0)}_{g}|$,
as  functions  of  the  momentum  component  $p_x$  and  the angle
$\theta$  between  $\vc{k}_{1}$  and  $\vc{k}_{2}$.  The Cartesian
coordinate    system    is introduced
\begin{figure}
\epsfbox{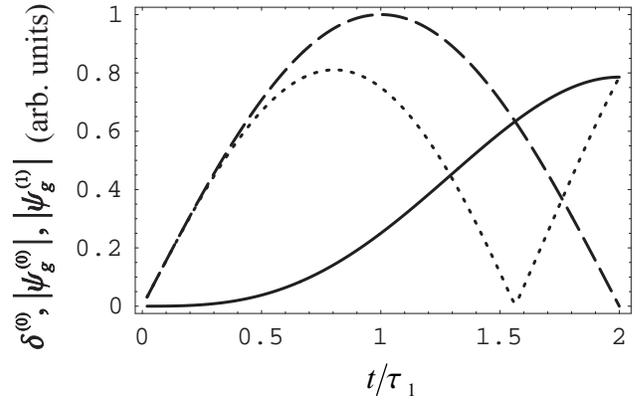}
\vspace{0.5cm}
\caption[Figure 4]{
    Time    dependences    of    $\delta^{(0)}$    (solid   line),
$|\psi^{(0)}_{g}|$  (long  dashed  line),  and  $|\psi^{(1)}_{g}|$
(short   dashed   line)   for   a   two-mode   object   beam  with
$\vc{k}_{1,2}\perp  \vc{k}_0$.    The  components  of $\vc{p}$ are
fixed  as  follows:    $p_y=  \hbar  k_0$  and $p_x= p_z= 0$.  The
coordinate axes and other parameters are the same as in Fig.\
\ref{fig3}.
}
\label{fig4}
\end{figure}
in   momentum   space,
$\vc{p}=(p_x,p_y,p_z)$,  with  the $x$   ($y$)  axis  directed   along
(opposite)  the  vector   $\vc{k}_{0}$  ($\vc{k}_{1}$).     Figure
\ref{fig3} shows the corresponding dependences after $\pi$-pulse  time
calculated for sodium atoms, provided that the initial wave packet
has the Gaussian profile $\psi_{g}(\vc{p}) \propto  \exp\left[-L^2
(\vc{p} - \vc{p}_{0})^2 /  (2 \hbar^2)\right]$ with mean  momentum
$\vc{p}_{0}=\hbar\vc{k}_{0}$,  $|\vc{k}_{0}|  =k_0  =  1.07 \times
10^5$ cm$^{-1}$, and spatial extension  $2L=0.4$ cm.  The peaks  in
the central region of each plot correspond to forbidden values  of
$\theta \propto \delta_k < \delta_p(\tau_1)$.  Outside these peaks
($|\theta|  >   10^{-4}$)  the   relative  correction   goes  down,
approaching $0.15$ at large $\theta$, that is, below its estimation
value $\varepsilon_{r} \approx 0.5$.

    In the worst case,  e.g., $\vc{k}_m \perp \vc{k}_0$,  $\forall
m\geq  1$,  the  estimation  for  $\varepsilon_{r}$ is higher than
$\delta_p(\tau_1)/(\hbar\delta_k)$,   because   it   depends    on
$\delta_{\omega}  =  \hbar  \delta_k^2/(2M)$,  the minimal kinetic
energy an atom can get due to transition between the laser modes:
\begin{equation}
\varepsilon_{r} \lesssim g_{\Sigma}\left|
\int\limits^{t}_{0}\sin^2(g_{\Sigma}\tau)
\exp(i \delta_{\omega}\tau) d\tau\right|.
\label{2.4.7}
\end{equation}
    Nevertheless, the term  $\chi(\vc{p},t)$ may still  be treated
as a small  perturbation if the  overall effective Rabi  frequency
satisfies the condition, more restrictive than Eq.\ (\ref{2.4.5}),
\begin{equation}
g_{\Sigma} \ll \delta_{\omega}.
\label{2.4.8}
\end{equation}
    Otherwise,  if   $g_{\Sigma}  \gtrsim   \delta_{\omega}$,  the
interaction time should be limited so that $t\ll \tau_1$.   Figure
\ref{fig4}    shows    time    dependences    of   $\delta^{(0)}$,
$|\psi^{(0)}_{g}|$,  and  $|\psi^{(1)}_{g}|$  for  a two-mode object
beam with $\delta_p(\tau_1)/(\hbar\delta_k)=0.1$ and  $g_{\Sigma}=
74 \delta_{\omega}$.   We see that  in the considered  unfavorable
configuration  $\varepsilon_{r}$  does  not  exceed  $0.1$ even if
$t=0.5\tau_1$.

    In most  practical  cases, however,  only a  small fraction  of
laser modes has a geometry leading to the condition (\ref{2.4.8}),
and the requirement (\ref{2.4.5}) appears to be sufficient.

\section{ATOM-OPTICS INTERPRETATION}
\label{sec3}

\subsection{General consideration}
\label{sec3.1}

    In an idealized situation, one may imagine that all atoms  are
initially in a pure state described by the Gaussian profile
\begin{equation}
g(\vc{p};\vc{p}_{0})=
\frac{L^{3/2}}{\hbar^{3/2}\pi^{3/4}}
\exp\left(\frac{-L^2 (\vc{p} - \vc{p}_{0})^2}
{2 \hbar^2} - \frac{i}{\hbar} \vc{p}\cdot\vc{r}_0\right)
\label{3.1.1}
\end{equation}
    with mean momentum $\vc{p}_0$ close to  $\hbar\vc{k}_0$,
space  position  $\vc{r}_0$,  and  very  small  dispersion [$L \gg
\hbar/ \delta_p(\tau_1)$].  According to Eq.\ (\ref{2.2.3}), after
interaction with the laser beams over a time period $\tau \lesssim
\tau_1$ and subsequent free propagation during time $t$ the  atoms
remain in a pure state, and their wave function can be represented
as       a       superposition       of       useful       signals
$\psi^{(s,r)}(\vc{p},\tau,t;\vc{p}_0)$     and     a    background
$\psi^{(b)}(\vc{p},\tau,t;\vc{p}_0)$, where
\begin{equation}
\psi^{(\sigma)}(\vc{p},\tau,t;\vc{p}_0) =
e^{-i w(\vc{p}) t}
\int d \vc{p}' G^{(\sigma)}(\vc{p},\vc{p}',\tau)
g(\vc{p}';\vc{p}_{0}),
\label{3.1.2}
\end{equation}
    $\sigma   \in   \{s,r,b\}$.      The  functions  $G^{(\sigma)}
(\vc{p},\vc{p}',\tau)$ are defined by  Eqs.\ (\ref{b.2}).   In the
case  considered  here  they  have explicit analytical expressions
relying on the analogy with Eqs.\ (\ref{c.1}).  These  expressions
are exact  in the  limit $L  \to \infty$,  and $\bbox{\kappa}  \to
\vc{0}$,   where   $\bbox{\kappa}=   \vc{p}_0/\hbar-    \vc{k}_0$.
Omitting      the      irrelevant      common     phase     factor
$\exp[-i\vc{k}_0\cdot\vc{r}_0    -i     f_0    \tau     -(i/\hbar)
E_{|g\rangle}(0)\tau]$  we  can  readily  infer  that  the Fourier
transform of $\psi^{(r)}(\vc{p},\tau,t;\vc{p}_0)$,
\begin{equation}
\psi_{r}(\vc{r},\tau,t;\vc{p}_0) =
A_{r}(\hbar\bbox{\kappa},\tau)
\gamma_0(\vc{r},\bbox{\kappa})
e^{i \vc{k}_0 \cdot \vc{r}
-i w(\hbar\vc{k}_0) t},
\label{3.1.3}
\end{equation}
    propagates like the reference beam.  Otherwise, the  transform
of $\psi^{(s)}(\vc{p},\tau,t;\vc{p}_0)$,
\begin{eqnarray}
\psi_{s}(\vc{r},\tau,t;\vc{p}_0)=&&
A_{s}(\hbar\bbox{\kappa},\tau)
\nonumber\\
&&\times\sum_{m\geq  1}
\gamma_m(\vc{r},\bbox{\kappa})
\frac{g_m}{g_{\Sigma}}
e^{i \vc{k}_m \cdot \vc{r}
-i w(\hbar\vc{k}_m) t},
\label{3.1.4}
\end{eqnarray}
    generates a  matter wave,  which inherits  the amplitude and
phase   characteristics   of   the   object   beam   because  $g_m
\propto\vc{E}_{m}$ as follows  from Eq.\ (\ref{a.3.c})  and
the definition  of the  Rabi frequencies  $\Omega_{m}$.   The last
assertion also takes into account that all functions
\begin{equation}
\gamma_m(\vc{r},\bbox{\kappa})=
\frac{L^{3/2}}{\pi^{3/4}\sigma^3}
\exp\left(
-\frac{L^2(\vc{r}-\tilde{\vc{r}}_t^m)^2}
{2|\sigma|^4}
+i \delta_{\phi}(\vc{r}-\vc{r}_t^m)\right)
\label{3.1.5}
\end{equation}
    used in Eqs.\ (\ref{3.1.3}) and (\ref{3.1.4}) slowly depend on
$\vc{r}$  within  spatial   regions  $\sim  2|\sigma|^2/L$,   each
centered     around     the     point     $\tilde{\vc{r}}_t^m    =
\vc{r}_t^m+\hbar(t+\tau) \bbox{\kappa}/M$, where
\begin{equation}
\sigma=\sqrt{L^2 +i\hbar(t+\tau)/M},
\label{3.1.6}
\end{equation}
\begin{equation}
\vc{r}_t^m =\vc{r}_0
+\frac{\hbar(\vc{k}_0+\tilde{\vc{k}})}{2M}\tau
+\frac{\hbar\vc{k}_m}{M} t,
\label{3.1.7}
\end{equation}
    and $\tilde{\vc{k}}$ is some typical wave vector in the object
beam.  The small phase shifts $\delta_{\phi} (\vc{r}-\vc{r}_t^m)$
\begin{equation}
\delta_{\phi}(\vc{r})=
\frac{1}{2|\sigma|^4}
\left(
L^4\vc{r}\cdot\bbox{\kappa}
+\frac{\hbar(t+\tau)(\vc{r}^2
-L^4\bbox{\kappa}^2)}{M}
\right)
\label{3.1.8}
\end{equation}
    introduced by $\gamma_m(\vc{r},\bbox{\kappa})$ vanish at small
$\bbox{\kappa}$ and large $L$.

    If the overall effective Rabi frequency is chosen in agreement
with  the  results  of  Sec.\  \ref{sec2.4}, the background, which
itself represents  a first-order  correction to  the wave function
$\psi^{(0)}_{g}(\vc{p},\tau)$,  appears  to  be  small at any time
$\tau   \lesssim   \tau_1$.      Consequently,  since  the  states
(\ref{3.1.3}) and (\ref{3.1.4}) are spatially separated after  the
free propagation  period $t_{\rm  min}=2L M/(\hbar\Delta_k)$,  one
may  observe  a  matter  wave   $\psi_{s}(\vc{r},\tau,t;\vc{p}_0)$
cloning  the  object  beam  in  the  space-time  region ${\cal S}=
\{(\vc{r},t):\;|\vc{r}- \tilde{\vc{r}}_t^m| < L,\; \forall m  \geq
1;\; t> t_{\rm min}\}$, where  all atomic wave packets related  to
different modes of this beam still overlap each other.  It  should
be noted that  ${\cal S}\ne \emptyset$  only when the  observation
time is limited  by the value  $t_{\rm max}= L  M/[\hbar \tilde{k}
\sin(\theta_{\rm max}/2)]$, where $\theta_{\rm max}$ characterizes
the maximal divergence angle  of the object beam,  and $\tilde{k}=
|\tilde{\vc{k}}|$.  In  a given context,  the physical meaning  of
conditions   (\ref{2.4.5}) and (\ref{2.4.8})   consists   in    the
requirement that a more delicate mechanism (lower laser intensity)
has to  be used  in order  to restore  more detailed  information.
The conditions have a  counterpart in the  theory of optical
holograms [see, e.g., Eq.\ (4) in Ref.\ \cite{Sidorovich}] which,
in turn,  is responsible for the low intensity of noise in
the reconstructed wave.

    In a  more realistic  case we  may expect  the initial  atomic
state to be a statistical mixture described by the density matrix
\begin{equation}
\rho_{gg}(\vc{p}_1,\vc{p}_2,0) =
\int d \vc{p}' f(\vc{p}')
g(\vc{p}_1;\vc{p}') g^{*}(\vc{p}_2;\vc{p}'),
\label{3.1.9}
\end{equation}
    where $f(\vc{p})$  denotes a  momentum distribution  function.
If this function is  compatible with the condition  (\ref{2.4.6}),
one   can   readily   obtain   an   expression   for    $\rho_{gg}
(\vc{p}_1,\vc{p}_2,t)$ at any time.  In the region ${\cal S}$,  it
takes the following form in the coordinate representation
\begin{equation}
\rho_{gg}(\vc{r}_1,\vc{r}_2,\tau,t) =
\int d \vc{p}' f(\vc{p}')
\psi_{s}(\vc{r}_1,\tau,t;\vc{p}')
\psi_{s}^{*}(\vc{r}_2,\tau,t;\vc{p}').
\label{3.1.10}
\end{equation}
    Since $A_{s}(\vc{p},\tau)$ is a sharply peaked function having
a  width  $\delta_p(\tau)$  along  the vector $\vc{\Delta}_k$ [see
Eq.\  (\ref{2.4.2})],  the  integral  in  Eq.\  (\ref{3.1.10})  is
limited in this direction.  Let us assume that integration in  the
transverse  directions  is  also  restricted  to  a  small  domain
$\sim\delta_p(\tau)$  due   to  the   finite  spectral   width  of
$f(\vc{p})$.   Then analyzing  Eq.\ (\ref{3.1.10}) 
\begin{figure}
\epsfbox{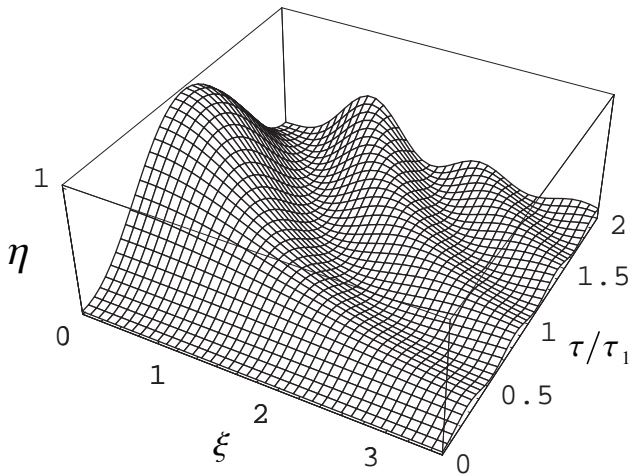}
\vspace{0.5cm}
\caption{
    Diffraction efficiency $\eta$ of atomic hologram as a function
of time domain $\tau$ (in units of $\pi$ pulse) and  dimensionless
parameter $\xi$.
}
\label{fig5}
\end{figure}
in the  region
$|\vc{r}_1-\vc{r}_2|\ll \hbar/\delta_p(\tau)$ one finds that under
the condition  $t\ll t_{\rm  coh} =  M/[\tilde{k} \sin(\theta_{\rm
max}/2)  \delta_p(\tau)]$  the  density  matrix  factorizes  as  a
product of coherent states,
\begin{equation}
\phi(\vc{r},\tau,t) =
C^{1/2}(\tau)\sum_{m\geq  1}
\gamma_m(\vc{r},\vc{0})
\frac{g_m}{g_{\Sigma}}
e^{i \vc{k}_m \cdot \vc{r}
-i w(\hbar\vc{k}_m) t},
\label{3.1.11}
\end{equation}
where
\begin{equation}
C(\tau) = \int d \vc{p}
\left|A_{s}(\vc{p},\tau)\right|^2
f(\vc{p}+\hbar\vc{k}_0).
\label{3.1.12}
\end{equation}
    When deriving  the formula  (\ref{3.1.11}) we  allow for small
values of phase differences $| \delta_{\phi} (\vc{r}_1-\vc{r}_t^m)
- \delta_{\phi} (\vc{r}_2-\vc{r}_t^n) |\ll \pi,\; \forall  m,n\geq
1$, appearing  in the  integration region  at $t\ll  t_{\rm coh}$,
whence  $\gamma_m(\vc{r},  \bbox{\kappa})  \simeq \gamma_m(\vc{r},
\vc{0})$.    Note  that  mutual  compatibility  of time conditions
$t_{\rm min} < t \ll t_{\rm coh}$ restricts the possible structure
of the object beam:
\begin{equation}
\sin(\theta_{\rm max}/2)\ll
\frac{\hbar\Delta_k}
{2 L \tilde{k}\delta_p(\tau)}.
\label{3.1.13}
\end{equation}

    We see that the superposition of laser beams selectively  acts
only on those  wave packets in  the initial representation  of the
atomic  density  matrix  Eq.\  (\ref{3.1.9})  whose  spectra  are
concentrated near  the vector  $\hbar\vc{k}_0$, and  thus restores
the pure state Eq.\ (\ref{3.1.11}).  Therefore the inhomogeneous  laser
radiation proves to behave like a three-dimensional hologram  with
respect to the incident atomic beam (impinging wave packets).

    One can further establish  a close relation between  an atomic
hologram created in a time domain $\tau$ and a permanent  optical
hologram with the  thickness $d_{\tau} =  \hbar k_0 \tau/M$  along
the  direction  of  the  reading  beam.   Indeed, as is known from
optics, the  passage of  the reading beam  through a three-dimensional
hologram can  be interpreted  as a  multiple diffraction  in which
small waves, diffracted  from different registration-medium  layers
with equivalent transmission of light, interfere constructively to
form a reconstructed wave of high intensity.  The same  approach
can  be  used  to  describe  an  atom-optics  hologram as a light
structure, inducing  an optical  potential through  the atom-laser
dipole   interaction   \cite{creation}.      Here   the   role  of
equivalent-transmission layers  in the  medium is  performed by the
equipotential surfaces.  Since $d_{\tau}$ is just the distance the
impinging wave packet  covers during time  $\tau$, the numbers  of
crossed interfaces (layers or  surfaces) are equal for  the atomic
and conventional holograms.  Therefore, if there were no  difference
in  the  initial  and  boundary  conditions, the processes of wave
front reconstruction would be identical in both cases.

    The relation  between atomic  and optical  holograms makes  it
possible to classify atomic holograms as thin or thick diffractive
optical elements, and use the Talbot length $L_{\rm Talbot}$, i.e.,
the  typical  interval   between  consecutive  interfaces,   as  a
characteristic  scale  to  distinguish  between  the  two  classes
\cite{Oberthaler}.  That is, the hologram can be considered as thick
(three dimensional) if $d_{\tau} > L_{\rm Talbot}$, or in terms of
time
\begin{equation}
\tau > L_{\rm Talbot} M/(\hbar k_0).
\label{3.1.14}
\end{equation}
    For most holographic  setups (for instance,  like that in
Fig.\ \ref{fig2})  $L_{\rm Talbot}\sim  2\pi/k_0$.   Therefore the
criterion (\ref{3.1.14})  leads to a time  domain $\tau$ larger
than  the  period  of  atomic  oscillations.  Obviously the latter
requirement is well satisfied for $\tau \sim \tau_1$, the time  of the
$\pi$ pulse, provided $g_{\Sigma}$ is chosen in agreement with the
condition (\ref{2.4.5}).

\subsection{Diffraction efficiency}
\label{sec3.2}

    In  a  regime  where  the  background  is small, we can define
the diffraction  efficiency  $\eta$  of  a  hologram  as  the  overall
intensity of the modes composing the reconstructed wave,  provided
the initial wave packet is normalized to $1$,
\begin{equation}
\eta(\tau,\vc{p}_0)=
\int d\vc{p} \left |
\psi^{(s)}(\vc{p},\tau,t;\vc{p}_0)
\right |^2.
\label{3.2.1}
\end{equation}
    It is clear, however, that $\eta$ depends on the shape of  the
initial distribution as well.  Therefore, to be more specific  let
us assume  a Gaussian  profile (\ref{3.1.1})  of the impinging  wave
packet  with  infinitely  small  dispersion  $L \to \infty$.  Then
integration  over  $\vc{p}'$  in  Eq.\  (\ref{3.1.2}) becomes
trivial, so that
\begin{equation}
\eta(\tau,\vc{p}_0)=
\int d\vc{p} \left |
G^{(s)}(\vc{p},\vc{p}_0,\tau)
\right |^2.
\label{3.2.2}
\end{equation}
    Using  the approximate  expressions  (\ref{c.1.b})  and   omitting
negligible interference terms one obtains from the above equation
\begin{equation}
\eta(\tau,\vc{p}_0)=
\eta(\tau,\xi)\simeq
\frac{1}{\xi^2+1}
\sin^{2}\left(
\tau g_{\Sigma} \sqrt{\xi^2+1}
\right),
\label{3.2.3}
\end{equation}
where the dimensionless parameter
\begin{equation}
\xi=\frac{(\vc{p}_{0}- \hbar\vc{k}_{0})
\cdot \vc{\Delta}_{k}}{M g_{\Sigma}}
\label{3.2.4}
\end{equation}
    characterizes  the  deviation  of  the initial atomic momentum
from the mean momentum of photons in the reference beam.

    According to  this simple  formula the  diffraction efficiency
achieves a maximum  at $\tau=\tau_n/  \sqrt{\xi^2+1}$ and  can reach
$100\%$ if $\xi= 0$ (see Fig.\ \ref{fig5}).

\subsection{Numerical example}
\label{sec3.3}

\begin{figure}
\epsfbox{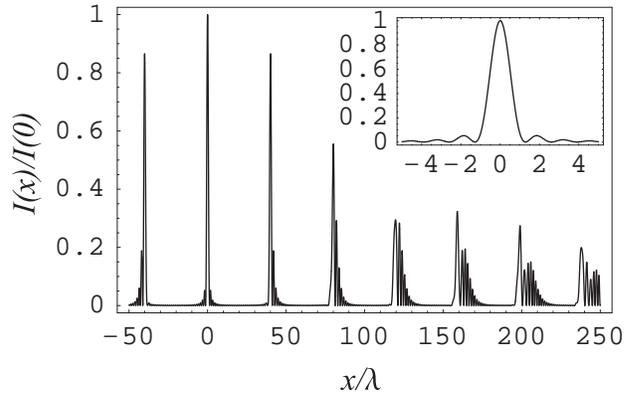}
\vspace{0.5cm}
\caption{
    Intensity $I(x)$ of the $31$-mode object wave as a function of
observation point $\vc{r}=(x,0,0)$.  The inset shows optical image
of a  single line  $\sim\lambda$  created in  the central  region
$\sim 60\lambda$.
}
\label{fig6}
\end{figure}

    In the following we show two-dimensional results obtained  for
Na  assuming  an  experimental  setup  like  that  in Fig.\
\ref{fig2}  (i.e.,  all  $\vc{k}_{m}$  are  coplanar  vectors  and
$\tilde{\vc{k}}\perp \vc{k}_{0}$).  The image to be  reconstructed
is a thin line of width $\lambda=2 \pi/k_{0}$ perpendicular to the
laser-beam plane.  To  decrease the amount of computational  work
we reduced  the number  of object  wave modes  to $31$  and set up
$\theta_{\rm max} = \pi/4$.   Such a field approaches the  desired
single  line  within  a  region  of size $\sim 60\lambda$ centered
around the point  $\vc{r}= \vc{0}$ if  all the laser  modes in the
expression (\ref{2.2.1})  have identical  amplitudes $\vc{E}_{m}$,
and their wave vectors $\vc{k}_{m}$ are equidistant
\begin{equation}
\vc{k}_{m}=k_0\left(
\sin\left[\frac{\pi(m-16)}{120} \right],\;
-\cos\left[\frac{\pi(m-16)}{120} \right],\;
0 \right).
\label{3.3.1}
\end{equation}
    The  corresponding  profile  of  the  object  beam   intensity
distribution $I(x)$ is  shown in Fig.\  \ref{fig6}, where the  $x$
axis  is  directed  along  the  vector  $\vc{k}_0=(k_0,0,0)$.  The
interference fringes, which are a consequence of the finite  number
of  modes,  can  easily be separated  from  the central line and
therefore do not contaminate our consideration.

    In numerical simulations the  optical pulse duration was  taken
to be $\tau_1= 2.82\times  10^{-2}$ s, to demonstrate  the highest
diffraction efficiency.  The remaining laser light parameters were
fixed  as  follows:    Rabi  frequencies  $\Omega_{0} = 1$ MHz and
$\Omega_{m} = 0.01 \Omega_{0}$ for all $1\leq m \leq 31$, detuning
$\Delta =  -1$ GHz  ($\gamma/\Delta \approx  0.06$), the effective
Rabi frequencies $g_m= 10$ Hz, $f_{mn}= 0.1$ Hz, and  $g_{\Sigma}=
55.7$  Hz.    Note  that  for  the considered laser-beam geometry
$\Delta_k  =  \sqrt{2}  k_{0}  =  1.51  \times 10^5$ cm$^{-1}$ and
$\delta_k= 3.28  \times 10^3$  cm$^{-1}$, so  that the  background
introduces  a relative  correction  of  the order $\varepsilon_{r} =
\delta_p(\tau_1)/ (\hbar\delta_k) = 1.4\times 10^{-2}$ and can  be
neglected.

\begin{figure}
\epsfbox{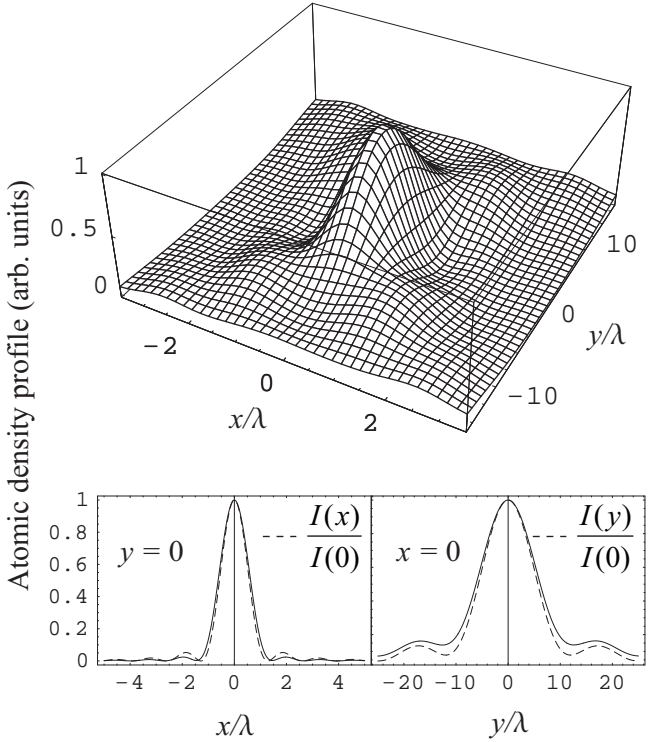}
\vspace{0.5cm}
\caption{
    Atomic  density  $\rho_{gg}  (\vc{r},\vc{r},\tau_1,t)$  as   a
function of observation point  $\vc{r}=(x,y,0)$.  The bottom  part
of the plot compares atomic profile (solid lines) with the  object
beam intensity distributions $I(x)$  and $I(y)$ (dashed lines)  in
the planes $y=0$ and $x=0$.
}
\label{fig7}
\end{figure}

    The reconstruction of a real image of the object was  achieved
by impinging  Gaussian wave  packets (\ref{3.1.1})  having spatial
extension $2L=0.4$ cm on the superposition of laser beams near the
point
\begin{equation}
\vc{r}_{0}=\left(
-\frac{\hbar k_0 \tau_1}{M},\;
\frac{L}{\tan(\theta_{\rm max}/2)}
-\frac{\hbar k_0 \tau_1}{M},\;
0 \right).
\label{3.3.2}
\end{equation}
    After the interaction with laser radiation is over, these wave
packets appear at a  distance $L/ \tan(\theta_{\rm max}/2)=  0.48$
cm from the image.  As  a result, the most intensive matter  field
in the imaging region may be observed after free propagation  time
$t= t_{\rm  max} \cos(\theta_{\rm  max}/2) =  0.16$ s,  which lies
within the limits $t_{\rm  min}= 9.6\times 10^{-2}$ s  and $t_{\rm
max}= 0.17$  s. Figure  \ref{fig7} shows  the corresponding atomic
density profile $\rho_{gg} (\vc{r},\vc{r},\tau_1,t)$ when the mean
momentum  of  the  initial   wave  packet  is  exactly   equal  to
$\hbar\vc{k}_0$.  As is seen from the bottom part of the  plot,
the atomic profile displays a good match with the distribution  of
the object  beam intensity.   The  attained diffraction efficiency
calculated according to Eq.\ (\ref{3.2.1}) is 98\% in this case.

\begin{figure}
\epsfbox{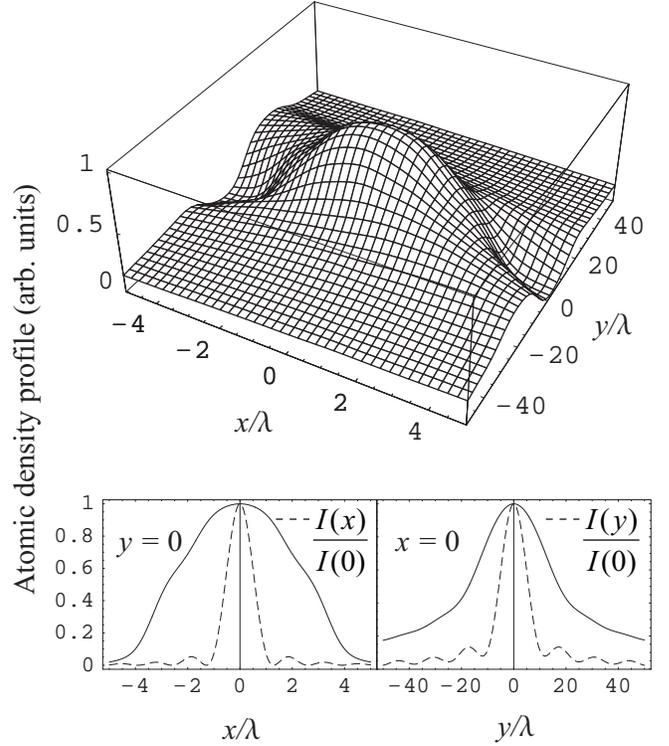}
\vspace{0.5cm}
\caption[Figure 8]{
    Atomic density  $\rho_{gg} (\vc{r},\vc{r},\tau_1,t)$  obtained
when  the  initial  state  is  a  statistical mixture with uniform
momentum distribution along the $x$ axis.  Other notations are the
same as in Fig.\ \ref{fig7}.
}
\label{fig8}
\end{figure}

    When initial state is a statistical mixture (\ref{3.1.9}) with
momentum distribution function  $f(\vc{p})$ uniform along  the $x$
axis, the atomic density  profile acquires a shape  represented in
the Fig.\  \ref{fig8}.   Since condition  (\ref{3.1.13}) does  not
hold  at  the  chosen  laser  light  parameters,  the  size of the
reconstructed line  appears to  be $\sim  4$ times  wider than one
might expect  from a  coherent reading  beam.   Nevertheless, such
image broadening is not too large, so that the atomic hologram can
be used even in this unfavorable design.

\section{CONCLUSIONS}
\label{sec4}

    In  this  paper  we  have  studied  a  method  of  driving
ultracold atom propagation using effective holograms made of laser
radiation in  a specified  time domain.   We  have shown  that the
scattered  atomic  wave  packet  may  inherit  the features of the
object electromagnetic  wave provided  the atomic  internal ground
state possesses  a translation  invariance due  to compensation of
gravity  with  the  Stern-Gerlach  effect.   We have established a
close relation 
between the  atomic hologram  created in  the time
domain and a thick optical hologram prepared in the  corresponding
spatial  region,  and  have  found  a  recipe for controlling the
diffraction efficiency of such an atomic hologram by means of varying
the time  domain.   Besides adjusting  the atom-laser  interaction
time, another way to enhance diffraction efficiency has proved  to
be the cooling of the atomic beam so that all the particles
get the same momentum as the momentum of photons in the  reference
wave.   A special  role here  may be  played by  BEC and  coherent
atomic-beam   generators,   which   are   under   development  now
\cite{CAB}.

    We have considered  dilute   atomic
samples,  i.e.,  we  have  not  included many-atom interactions
\cite{Lewenstein}, which may lead to nonlinear atom-optics effects
\cite{nonlinear}  along  with  enhancing  the  background.     The
conditions under  which these  interactions can  be neglected were
elaborated in our previous paper \cite{creation} using the mean-field
approximation applied to the Maxwell-Bloch equations \cite{Castin}
and are well satisfied when the mean-field interaction energy  per
particle is much less than the typical kinetic energy of an  atom.
We have also neglected such possible sources of background  as
spontaneous  emission  of  photons  and  fluctuations of the laser
frequency.  While the first of these sources may be eliminated  by
keeping  the  laser  
detuning  much  bigger  than  the spontaneous
emission rate, the second one is determined by the spectral  width
of    two-time    electromagnetic-field    correlation   functions
\cite{creation,Dalton}, and substantially  decreases if all  field
modes originate from one initial laser mode.

    Although our scheme of an atomic hologram has been developed  for
codirected reading and reference beams it can readily be modified
for an experimental setup with opposite propagation of the beams.
In full analogy with conventional optics, such a hologram will
reconstruct the conjugate object wave.

    In conclusion,  atom-optics  holograms appear  to be  a useful
implement for solving some of the basic technological problems  in
the field of  atom lithography.   For instance, they  will make it
possible to grow 3D  circuitry components by depositing an  arbitrary
multilayer pattern of impurity atoms on a silicon substrate.

\appendix
\section{Elimination of the excited state}
\label{appenda}

    In   the   rotating   wave   approximation  the  two-component
Schr\"{o}dinger  equation,  rewritten  for
ground- and excited-level wave functions $\psi_{g}(\vc{p},t)$  and
$\psi_{e}(\vc{p},t)$ slowly  varying in time, takes the form
\begin{mathletters}
\begin{eqnarray}
i\frac{\partial}{\partial t} \psi_{g}(\vc{p},t) = &&
[w(\vc{p})+\Delta]\psi_{g}(\vc{p},t) \nonumber\\
&&-\sum_{m\geq 0}
\Omega^{*}_{m} \psi_{e}(\vc{p}+\hbar\vc{k}_{m},t),
\label{a.1.a}
\end{eqnarray}
\begin{eqnarray}
i\frac{\partial}{\partial t} \psi_{e}(\vc{p},t) = &&
[w(\vc{p})-i\vc{f}_e\cdot\vc{\nabla}]\psi_{e}(\vc{p},t)
\nonumber\\
&&-\sum_{m\geq 0}
\Omega_{m} \psi_{g}(\vc{p}-\hbar\vc{k}_{m},t),
\label{a.1.b}
\end{eqnarray}
\label{a.1}
\end{mathletters}
    where $\Omega_{m}$ stands for  the Rabi frequency of  mode $m$
and    the     terms    $w(\vc{p})=     \vc{p}^2/(2M\hbar)$    and
$-i\vc{f}_e\cdot\vc{\nabla}$  arise  in  momentum  space  from the
kinetic    and    potential    energy     ($-\vc{f}_e\cdot\vc{r}$)
respectively.

    The  route  by  which  one  can  adiabatically  eliminate  the
component $\psi_{e}(\vc{p},t)$  from Eqs.\  (\ref{a.1}) implies  a
self-consistent assumption $|\psi_{e}| \ll |\psi_{g}|$ leading  to
the    zero-order    solution    of   Eq.\   (\ref{a.1.a}):
$\psi_{g}(\vc{p},t)  \simeq   \exp\{-i[w(\vc{p})  +   \Delta]  t\}
\psi_{g}(\vc{p},t=0)$.  After substitution of this expression into
Eq.\  (\ref{a.1.b})  the  latter  is  solved  perturbatively  with
respect to the potential energy term:
\begin{eqnarray}
\psi_{e}(\vc{p},t)\simeq &&
-\sum_{m\geq 0}
\frac{\Omega_{m} \psi_{g}(\vc{p}-\hbar\vc{k}_{m},t)}
{w(\vc{p})-w(\vc{p}-\hbar\vc{k}_{m})-\Delta}
+ \ldots,
\label{a.2}
\end{eqnarray}
    where  the  ellipsis  denotes  omitted  terms that include a small
($\propto |\vc{f}_e|L/|\hbar  \Delta|$) first-order  correction to
$\psi_{e}(\vc{p},t)$  and  also  terms  that  oscillate  with the
nonresonant frequency $w(\vc{p})$ and therefore give a negligible
contribution when one uses the above expression within the context
of Eq.\ (\ref{a.1.a}).

    For an  ultracold atomic  sample one  can further  discard the
kinetic  energy  terms  in  the  denominators  of  the  expression
(\ref{a.2}).   As a  result, Eq.\ (\ref{a.1.a})  takes the
form of Eq.\ (\ref{2.2.5}), provided  that the effective  Rabi frequencies
$f_0$, $f_{mn}$, and $g_{m}$ are defined as follows:
\begin{mathletters}
\begin{equation}
f_0=\frac{1}{\Delta}
\sum_{m\geq 0}|\Omega_{m}|^2,
\label{a.3.a}
\end{equation}
\begin{equation}
f_{mn}=\frac{\Omega_{m} \Omega^{*}_{n}}{\Delta},
\label{a.3.b}
\end{equation}
\begin{equation}
g_{m}=\frac{\Omega_{m} \Omega^{*}_{0}}{\Delta}.
\label{a.3.c}
\end{equation}
\label{a.3}
\end{mathletters}

\section{Ground-state Green function}
\label{appendb}

    Here  we   present  the   first-order  approximation   to  the
ground-state component of the Green function determining the  time
evolution  of  the  atomic  density  matrix  according  to formula
(\ref{2.2.3}):
\begin{equation}
G_{gg}(\vc{p},\vc{p}',t) = e(t)
\sum_{\sigma \in \{r,s,b\}}
G^{(\sigma)}(\vc{p},\vc{p}',t),
\label{b.1}
\end{equation}
    where  common   phase  multiplier   $e(t)  =   \exp[i\omega  t
-(i/\hbar)    E_{|e\rangle}(0)t]$     recovers    the     solution
(\ref{2.3.12}) from its slow time dependence, and
\begin{mathletters}
\begin{equation}
G^{(r)}(\vc{p},\vc{p}',t) =
{\cal M}\left[
\phi^{(0)}_{0}(\vc{p},
\vc{p}',\lambda)
\right],
\label{b.2.a}
\end{equation}
\begin{equation}
G^{(s)}(\vc{p},\vc{p}',t) =
\sum_{m \geq 1}
{\cal M}\left[
\phi^{(0)}_{m}(\vc{p},
\vc{p}',\lambda)
\right],
\label{b.2.b}
\end{equation}
\begin{equation}
G^{(b)}(\vc{p},\vc{p}',t) =
\sum_{m \geq 0}
{\cal M}\left[
\phi^{(b)}_{m}(\vc{p},
\vc{p}',\lambda)
\right].
\label{b.2.c}
\end{equation}
\label{b.2}
\end{mathletters}

    In these formulas the  operator ${\cal M}$ stands  for the inverse
Laplace transformation and shift of the momentum arguments
\begin{eqnarray}
{\cal M}\left[
\phi^{(\sigma)}_{m}(\vc{p},\vc{p}',\lambda)
\right] \equiv && 2\pi i
\int_{\epsilon-i\infty}^{\epsilon+i\infty}
d\lambda  e^{\lambda t}
\nonumber\\
&&\times
\phi^{(\sigma)}_{m}(\vc{p}-\hbar\vc{k}_m,
\vc{p}',\lambda),
\label{b.3}
\end{eqnarray}
$\epsilon >0$, $\sigma \in \{0,b\}$, $m\geq 0$, whereas
\begin{mathletters}
\begin{equation}
\phi^{(0)}_{0}(\vc{p},\vc{p}',\lambda) =
\frac{-i}{T(\vc{p},\lambda)}
\delta^3(\vc{p}+\hbar\vc{k}_0-\vc{p}'),
\label{b.4.a}
\end{equation}
\begin{equation}
\phi^{(b)}_{0}(\vc{p},\vc{p}',\lambda) =
\frac{i}{T(\vc{p},\lambda)^2}
\chi(\vc{p},\vc{p}',\lambda),
\label{b.4.b}
\end{equation}
\begin{equation}
\phi^{(\sigma)}_{m}(\vc{p},\vc{p}',\lambda) =
\frac{-g_m}{w_m(\vc{p}) -i\lambda}
\phi^{(\sigma)}_{0}(\vc{p},\vc{p}',\lambda),
\label{b.4.c}
\end{equation}
\label{b.4}
\end{mathletters}
    $m\geq       1$,       and       the       expression      for
$\chi(\vc{p},\vc{p}',\lambda)$ is obtained from Eq.\ (\ref{2.3.7}),
\begin{eqnarray}
\chi(\vc{p},\vc{p}',\lambda) = &&
\sum_{m\geq 1}\biggl(
\sum_{\scriptstyle n\geq 0 \atop \scriptstyle n\neq m}
g^{*}_{m} \phi^{(b)}_{n}[\vc{p}
+\hbar(\vc{k}_{m}-\vc{k}_{n}),\vc{p}',\lambda]
\nonumber\\
&&+\sum_{n\geq 1}
g_{n} \phi^{(b)}_{m}[\vc{p}+
\hbar(2\vc{k}_{0}-\vc{k}_{n}-\vc{k}_{m}),\vc{p}',\lambda]
\nonumber\\
&&+\sum_{\scriptstyle n\geq 1 \atop \scriptstyle n\neq m}
\sum_{l\geq 0}  f_{mn}
\nonumber\\
&&\times \phi^{(b)}_{l}[\vc{p}
+\hbar(\vc{k}_{0}-\vc{k}_{l}-\vc{k}_{m}+\vc{k}_{n}),
\vc{p}',\lambda]
\biggr).
\nonumber\\
\label{b.5}
\end{eqnarray}

\section{Near-resonance approximation}
\label{appendc}

    Here we develop the  near-resonance approximation in order  to
get  an  explicit  solution  of  the truncated system of equations
(\ref{2.3.4}) and (\ref{2.3.6}).   In the region  ${\cal D}$ one  can
approximately   treat   the   kinetic energy  terms  $w_m(\vc{p})$
corresponding to the  different modes of  the object beam  ($m\geq
1$)   as   equal:      $w_m(\vc{p})  \approx  w_n(\vc{p})  \approx
\tilde{w}(\vc{p})$,        where        $\tilde{w}(\vc{p})       =
w(\vc{p}+\hbar\tilde{\vc{k}}) +\Delta + f_0$, and $\tilde{\vc{k}}$
is  some  typical  wave  vector  in  the  object beam.  Under this
condition  the  integral  in  Eq.\  (\ref{2.3.12})  can   be
calculated     explicitly,     and     the     wave      functions
$\psi^{(0)}_{m}(\vc{p},t)$    acquire    a    simple    analytical
representation,
\begin{mathletters}
\begin{equation}
\psi^{(0)}_{0}(\vc{p},t) \simeq
A_{r}(\vc{p},t)
e^{-i b(\vc{p}) t}
\psi_{g}(\vc{p}+\hbar\vc{k}_{0}),
\label{c.1.a}
\end{equation}
\begin{equation}
\psi^{(0)}_{m}(\vc{p},t) \simeq
A_{s}(\vc{p},t)
\frac{g_{m}} {g_{\Sigma}}
e^{-i b(\vc{p}) t}
\psi_{g}(\vc{p}+\hbar\vc{k}_{0}),
\quad m\geq 1.
\label{c.1.b}
\end{equation}
\label{c.1}
\end{mathletters}
In these formulas,
\begin{mathletters}
\begin{equation}
A_{r}(\vc{p},t)=
\frac{i a(\vc{p})} {d(\vc{p})}
\sin[d(\vc{p}) t]
+ \cos[d(\vc{p})t],
\label{c.2.a}
\end{equation}
\begin{equation}
A_{s}(\vc{p},t)=
\frac{-i g_{\Sigma} } {d(\vc{p})}
\sin[d(\vc{p}) t],
\label{c.2.b}
\end{equation}
\label{c.2}
\end{mathletters}
where
\begin{equation}
a(\vc{p})= [\tilde{w}(\vc{p})
- w_{0}(\vc{p})]/2,
\label{c.3}
\end{equation}
\begin{equation}
b(\vc{p}) =
a(\vc{p}) + w_{0}(\vc{p}),
\label{c.4}
\end{equation}
\begin{equation}
d(\vc{p})= \sqrt{a(\vc{p})^2
+ g_{\Sigma}^2},
\label{c.5}
\end{equation}
and
\begin{equation}
g_{\Sigma}=\left(
\sum_{m\geq 1} |g_m|^2
\right)^{1/2}
\label{c.6}
\end{equation}
stands for the overall effective Rabi frequency.

%


\begin{references}
%
\bibitem{VSCPT}A. Aspect, E. Arimondo, R. Kaiser, N.
Vansteenkiste, and C. Cohen-Tannoudji,
Phys. Rev. Lett. {\bf 61}, 826 (1988); J. Lawall, S. Kulin, B.
Saubamea, N. Bigelow, M. Leduc, and C. Cohen-Tannoudji, {\it
ibid}. {\bf 75}, 4194 (1995).
%
\bibitem{Raman}M. Kasevich and S. Chu, Phys. Rev. Lett. {\bf
69}, 1741 (1992); N. Davidson, H. J. Lee, M. Kasevich, and S. Chu,
{\it ibid}. {\bf 72}, 3158 (1994); H. J. Lee, C. S. Adams, M.
Kasevich, and S. Chu, {\it ibid}. {\bf 76}, 2658 (1996).
%
\bibitem{amirror}V. I. Balykin, V. S. Letokhov, Yu. B.
Ovchinnikov, and A. I. Sidorov, Pis'ma Zh. Eksp.
Teor. Fiz. {\bf 45}, 282 (1987) [JETP Lett. {\bf 45}, 353 (1987)];
M. Arndt, P. Szriftgiser, J. Dalibard, and A. M. Steane,
Phys. Rev. A {\bf 53}, 3369 (1996);
N. Friedman, R. Ozeri, and N. Davidson, J. Opt. Soc. Am. B
{\bf 16}, 1749 (1998).
%
\bibitem{Moskowitz}P. E. Moskowitz, P. L. Gould, S. R. Atlas, and
D. E. Pritchard, Phys. Rev. Lett. {\bf 51}, 370 (1983).
%
\bibitem{Martin}P. J. Martin, B. G. Oldaker, A. N. Miklich, and
D. E. Pritchard, Phys. Rev. Lett. {\bf 60}, 515 (1988).
%
\bibitem{alens}V. I. Balykin, I. I. Klimov, and V. S. Letokhov,
Pis'ma Zh. Eksp. Teor. Fiz. {\bf 59}, 219 (1994) [JETP Lett. {\bf 59}, 235 (1994)];
M. K. Olsen, T. Wong, S. M. Tan, and D. F. Walls, Phys. Rev. A
{\bf 53}, 3358 (1996).
%
\bibitem{Gabor}D. Gabor, Nature (London) {\bf 161}, 777 (1948);
Proc. R. Soc. London, Ser. A {\bf 197}, 454 (1949).
%
\bibitem{atomlithography}G. Timp, R. E. Behringer, D. M. Tennant,
J. E. Cunningham, M. Prentiss, and K. K. Berggren,
Phys. Rev. Lett. {\bf 69}, 1636 (1992); J. J. McClelland,
R. E. Scholten, E. C. Palm, and R. J. Celotta, Science
{\bf 262}, 877 (1993); R. Gupta, J. J. McClelland, Z. J. Jabbour,
and R. J. Celotta, Appl. Phys. Lett. {\bf 67}, 1378 (1995).
%
\bibitem{Moringa}M. Moringa, M. Yasuda, T. Kishimoto, F.
Shimizu, J.T. Fujita, and S. Matsui, Phys. Rev. Lett. {\bf 77}, 802 (1996).
%
\bibitem{ahbose}O. Zobay, E. V. Goldstein, and P. Meystre,
Phys. Rev. A {\bf 60}, 3999  (1999).
%
\bibitem{BEC}M. Anderson, J. R. Ensher, M. R. Matthews, C. E. Wieman,
and E. A. Cornell, Science {\bf 269}, 198 (1995);
K. B. Davis, M.-O. Mewes, M. R. Andrews, N. J. van Druten,
D. S. Durfee, D. M. Kurn, and W. Ketterle, Phys. Rev. Lett.
{\bf 75}, 3969 (1995);
C. C. Bradley, C. A. Sackett, J. J. Tollett, and R. Hulet, {\it ibid}. {\bf 75}, 1687 (1995).
%
\bibitem{creation}A. V. Soroko, J. Phys. B. {\bf 30}, 5621 (1997).
%
\bibitem{Olshanii}M. Olshanii, N. Dekker, C. Herzog, and M. Prentiss,
e-print quant-ph/9811021.
%
\bibitem{Kogelnik}H. Kogelnik, Bell Syst. Techn. J. {\bf 48},
2909 (1969).
%
\bibitem{Ewald}P. P. Ewald, Ann. Phys. (Leipzig) {\bf 54}, 519
(1917).
%
\bibitem{Sidorovich}V. G. Sidorovich, Zh. Tekh. Fiz. {\bf 46},
1306 (1976) [Sov. Phys. Tech. Phys. {\bf 21}, 742 (1976)].
%
\bibitem{Oberthaler}M. K. Oberthaler, R. Abfalterer, S. Bernet,
C. Keller, J. Schmiedmayer, and A. Zeilinger, Phys. Rev. A {\bf 60},
456 (1999).
%
\bibitem{cooling}A. V. Soroko, Phys. Rev. A {\bf 58}, 3963 (1998).
%
\bibitem{alaser}M.-O. Mewes, M. R. Andrews, D. M. Kurn, D. S. Durfee,
C. G. Townsend, and W. Ketterle, Phys. Rev. Lett. {\bf 78},
582 (1997).
%
\bibitem{Moler}K. Moler, D. S. Weiss, M. Kasevich, and S. Chu,
Phys. Rev. A {\bf 45}, 342 (1992).
%
\bibitem{Korsunsky}E. A. Korsunsky, D. V. Kosachiov, B. G.
Matisov, and Yu. V. Rozhdestvensky, Zh. Eksp. Teor. Fiz.
{\bf 103}, 396 (1993)
[JETP {\bf 76}, 210 (1993)].
%
\bibitem{Kazantsev}A. P. Kazantsev, G. A. Ryabenko, G. I.
Surdutovich, and V. P. Yakovlev, Phys. Rep. {\bf129}, 75 (1985).
%
\bibitem{CAB}R. J. C. Spreeuw, T. Pfau, U. Janicke, and M.
Wilkens, Europhys. Lett. {\bf 32}, 469 (1995); H. M. Wiseman and
M. J. Collett, Phys. Lett. A {\bf 202}, 246 (1995);
M. Holland, K. Burnett, C. Gardiner, J. I. Cirac, and P. Zoller,
Phys. Rev. A {\bf 54}, R1757 (1994);
A. M. Guzman, M. Moore, and P. Meystre, {\it ibid}. {\bf 53},
977 (1996); G. M. Moy, J. J. Hope, and C. M. Savage, {\it ibid}.
{\bf 55}, 3631 (1997).
%
\bibitem{Lewenstein}M. Lewenstein, L. You, J. Cooper, and
K. Burnett, Phys. Rev. A {\bf 50}, 2207 (1994).
%
\bibitem{nonlinear}W. Zhang and D. F. Walls, Phys. Rev. A {\bf 49},
3799 (1994);  G. Lenz,  P. Meystre,  and E.  M. Wright,
{\it ibid}. {\bf 50}, 1681 (1994).
%
\bibitem{Castin}Y. Castin and K. M{\o}lmer,
Phys.  Rev.  A {\bf 51}, R3426 (1995).
%
\bibitem{Dalton}B. J. Dalton and P. L. Knight, J. Phys. B. {\bf
15}, 3997 (1982).
%
\end{references}
\end{document}